\documentclass[
    ,final            
  ]
  {aipproc}

\layoutstyle{8x11single}

\usepackage[latin1]{inputenc}
\usepackage[english]{babel}
\usepackage{graphicx}
\usepackage{lscape}
\usepackage{makeidx}
\usepackage{indentfirst}

\newcommand{\A}{{\!A\vphantom{\overline{J}}}}
\newcommand{\B}{{\!B\vphantom{\overline{J}}}}
\newcommand{\sqA}{{\sqrt{\rho_\A}}}
\newcommand{\sqB}{{\sqrt{\rho_\B}}}
\newcommand{\trac}{{\rm Tr}}
\newcommand{\Tr}{{\rm Tr}}
\newcommand{\diag}{{\rm diag}}
\newcommand{\Otimes}{{ \otimes }}
\newcommand{\Boltz}{ k_{\rm\scriptscriptstyle B}}
\newcommand{\Ran}{{\rm Ran}}
\newcommand{\ddt}[1]{{\frac{{\rm d}#1}{{\rm d}t}}}

\newcommand{\be}{\begin{equation}}
\newcommand{\ee}{\end{equation}}
\newcommand{\lt}{\left(}
\newcommand{\rt}{\right)}
\newcommand{\lgr}{\left\{}
\newcommand{\rgr}{\right\}}
\newcommand{\fr}{\frac}
\newcommand{\lqu}{\left[}
\newcommand{\rqu}{\right]}

\begin{document}

\title{NONLINEAR DYNAMICAL EQUATION FOR IRREVERSIBLE,
STEEPEST-ENTROPY-ASCENT RELAXATION TO STABLE EQUILIBRIUM}

\classification{03.65.Ta, 11.10.Lm, 04.60.-m, 05.45.-a}

\keywords {Nonlinear Dynamics Dynamics; Irreversibility;  Entropy;
 Information Theory;  Quantum Thermodynamics}

\author{Gian Paolo Beretta}{
  address={Universit\`a di Brescia, Italy}
}

\begin{abstract}
We discuss the structure and  main features of the nonlinear
evolution equation proposed by this author as the fundamental
dynamical law within the framework of Quantum Thermodynamics. The
nonlinear equation generates a dynamical group providing a unique
deterministic description of irreversible, conservative relaxation
towards equilibrium from any non-equilibrium state, and satisfies
a very restrictive stability requirement equivalent to
Hatsopoulos-Keenan statement of the second law of thermodynamics.
Here, we emphasize its mathematical structure and its
applicability also within other contexts, such as Classical and
Quantum Statistical Mechanics, and Information Theory.
\end{abstract}

\maketitle

\section*{INTRODUCTION}

The problem of understanding entropy and irreversibility has been
tackled by a large number of preeminent scientists during the past
century. Schools of thought have formed and flourished around
different perspectives of the problem. But a definitive solution
has yet to be found.

We address a mathematical problem very relevant to the question of
nonequilibrium and irreversibility, namely, that of ``designing''
a general evolution equation capable of describing irreversible
but conservative relaxation towards equilibrium. Our objective is
to present an interesting mathematical solution to this ``design''
problem, namely, a new nonlinear evolution equation that satisfies
a set of very stringent, relevant requirements \cite{MPLA1}.

We first define four essentially different contexts within which
the new equation is clearly relevant, with entirely different
interpretations:  Classical Statistical Mechanics (CSM), Classical
Information Theory (CIT), Quantum Statistical Mechanics (QSM),
Quantum Thermodynamics (QT). Then, we list the ``design
specifications'' that we intend to impose on the desired evolution
equation. We review some useful well-known mathematics involving
Gram determinants and, finally, present our nonlinear evolution
equation which meets the stringent design specifications.

 The discussion here is purposely devoided of our school's quite unorthodox
perspective, views and hypotheses on the physical meaning of
entropy and irreversibility. This is not only due to obvious space
limitations, but mainly because we feel that the proposed
nonlinear equation constitutes an important mathematical-physics
advance in itself not only in the Quantum Thermodynamics contexts
for which  it was originally designed and developed
\cite{HG,thesis,Frontiers,Cimento1,Cimento2,TwoLevel}, but also in
other contexts, such as CSM, CIT, QSM, as well as possibly in
Quantum Information, Biology, Sociology and Economics.

\section{FRAMEWORK A: CLASSICAL STATISTICAL MECHANICS}

Let $ \Omega $ be a phase space, and $ \mathcal{L} $ the set of
real, square-integrable functions $ A , B , \ldots $ on $ \Omega
$, equipped with the inner product $ ( \cdot | \cdot ) $ defined
by \be (A | B) = \trac (A B) = {\textstyle \int_{\Omega}} A B \,\,
d \Omega \label{1b} \ee \noindent where $ \trac (\cdot) $ in this
framework denotes $ \int_{\Omega} \cdot\,d \Omega $. We denote by
$ \mathcal{P} $ the subset of all nonnegative-definite, normalized
functions (distributions) $ \rho $ in $ \mathcal{L} $, i.e., \be
\mathcal{P} = \big\{ \rho \mbox{ in } \mathcal{L}\, |\, \rho \geq
0 , \trac ( \rho) = {\textstyle \int_{\Omega} } \rho \, d \Omega =
1 \big\} \label{2b} \ee \noindent We will then consider a set $
\lgr H , N_1 , \ldots , N_r \rgr $ of functions in $ \mathcal{L}
$.

In Classical Statistical Mechanics, $ \rho $ is the Gibbs
density-of-phase distribution function which represents the index
of statistics from a generally heterogeneous ensemble of identical
systems (with associated phase space $ \Omega $) distributed over
a range of possible classical mechanical states (the support of
$\rho$). $ H $ is the Hamiltonian function, and $ N_i $ the
number-of-particle function for particles of type $ i $.

\section*{FRAMEWORK B: CLASSICAL INFORMATION THEORY}

Let $ \mathcal{L} $ be the set of all $ n \times n $ real,
diagonal matrixes $ A = \diag ( a_j )$, $B = \diag ( b_j )$, \dots
( $n \leq \infty$ ), equipped with the inner product $ ( \cdot |
\cdot ) $ defined by \be \lt A | B \rt = \trac (A B) = {\textstyle
\sum_{j = 1}^n} a_j\, b_j \label{1c} \ee \noindent We denote by $
\mathcal{P} $ the subset of all nonnegative-definite, unit-trace
matrixes $ \rho $ in $ \mathcal{L} $, i.e.,

\be \mathcal{P} = \big\{ \rho = \diag ( p_j )\, |\, p_j \geq 0 ,\
\trac ( \rho) = {\textstyle \sum_{j = 1}^n} p_j = 1 \big\}
\label{2c} \ee \noindent Later we consider a set $ \lgr H , N_1 ,
\ldots , N_r \rgr    $  of diagonal matrixes $ H = \diag ( e_j )$,
$N_1 = \diag ( n_{1 j } )$, \dots, $N_r = \diag ( n_{rj} ) $ in $
\mathcal{L} $.

In Information Theory \cite{Jaynes}, $ \rho = \diag ( p_j ) $
represents the probability assignment to a set of $ n $ events, $
p_j $ being the probability of occurrence of the $ j $-th event. $
H $, $N_1 $, \dots, $N_r $ are characteristic features of the
events in the set, taking on the values $ e_j $, $n_{1 j} $, \dots
, $n_{r j} $, respectively, for the $ j $-th event.

\section*{FRAMEWORK C: QUANTUM STATISTICAL MECHANICS}

Let $ \mathcal{H} $ be a Hilbert space (dim $ \mathcal{H} \leq
\infty $), and $ \mathcal{L} $ the set of all linear operators $ A
$, $B $, \dots  on $ \mathcal{H} $, equipped with the real inner
product $( \cdot | \cdot )$ defined by \be \lt A | B \rt = \trac
\lt A^{\dag} B + B^{\dag} A \rt/2 \label{1a} \ee \noindent where $
A^{\dag} $ denotes the adjoint of operator $ A $ and $ \trac(
\cdot ) $ the trace functional. We denote by $ \mathcal{P} $ the
set of all self-adjoint, nonnegative-definite, unit-trace
operators $ \rho $ in $ \mathcal{L} $, i.e., \be \mathcal{P} =
\lgr \rho \mbox{ in } \mathcal{L} | \rho^{\dag} = \rho , \rho \geq
0 , \trac \rho = 1 \rgr \label{2a} \ee \noindent We will then
consider a set $ \lgr H , N_1 , \ldots , N_r \rgr $ of
self-adjoint operators in $ \mathcal{L} $, where each $ N_i $
commutes with $ H $, i.e., is such that $ H N_i = N_i H $, for $ i
= 1 , \ldots , r $.

In Quantum Statistical Mechanics, $ \rho $ is the von Neumann
statistical or density operator which represents the index of
statistics from a generally heterogeneous ensemble of identical
systems (same Hilbert space $ \mathcal{H} $ and operators $\lgr H
, N_1 , \ldots , N_r \rgr$) distributed over a range of generally
different quantum mechanical  states. If each individual member of
the ensemble is isolated and uncorrelated from the rest of the
universe, its state is described according to Quantum Mechanics by
an idempotent density operator
($\rho^2=\rho=P_{|\psi\rangle}=\frac{|\psi\rangle\langle\psi|}{\langle\psi|\psi\rangle}$),
i.e., a projection operator onto the span of some vector
$|\psi\rangle$ in $ \mathcal{H} $. If the ensemble is
heterogeneous, its individual member systems may be in different
states, $P_{|\psi_1\rangle}$, $P_{|\psi_2\rangle}$, and so on.

$ H $ is the Hamiltonian operator, and operator $ N_i $, for $ i =
1$, \dots , $r $, is the number operator for particles of type $ i
$ in the system (if the system has a fixed number $ n_i $ of
particles of type $ i $, then $ N_i = n_i I $, where $ I $ is the
identity operator on $ \mathcal{H} $).

\section*{FRAMEWORK D: QUANTUM THERMODYNAMICS}

In our formulation of Quantum Thermodynamics
\cite{HG,thesis,Frontiers,Cimento1, Cimento2}, the mathematical
framework is the same as that just summarized for QSM, but the
fundamental difference is in the physical meaning of the density
operator. Indeed, QT assumes that  the true individual quantum
state of a system isolated and uncorrelated from the rest of the
universe is represented by a density operator $\rho$ which are not
necessarily idempotent. Over the set of idempotent $\rho$'s, QT
coincides with Quantum Mechanics, but it differs fundamentally
from it because it assumes a broader set of possible states,
corresponding to the set of non-idempotent $\rho$'s. This way, the
functional $S(\rho)$ (defined in the next section) describes in QT
an intrinsic (non-statistical) fundamental state property of the
individual system. This is very different from the meaning that
the von Neumann  functional $S(\rho)$ has in QSM, where it
measures the degree of heterogeneity of the ensemble whose
statistics are represented by $\rho$.

\section*{MEAN VALUE FUNCTIONALS AND $ S $-FUNCTIONAL}

From here on, our notation allows us to treat at once the four
contexts just defined. For reasons to become apparent below,  the
elements $ H $, $N_1 $,\dots, $N_r $ introduced in either context,
will be called the generators of the motion (MG). We  assume that
such sets always contain at least element $ H $, that we call the
Hamiltonian MG.

For each MG, we then define a mean value functional on $
\mathcal{P} $ as follows

\be m ( \rho ; H ) = \trac \rho H = \lt \sqrt{\rho} | \sqrt{\rho}
H \rt \ \ ,\qquad m ( \rho ; N_i ) = \trac \rho N_i = \lt
\sqrt{\rho} | \sqrt{\rho} N_i \rt \label{4} \ee \noindent
Moreover, we define the $ S $-functional \cite{Lyapunov} on $
\mathcal{P} $ as \be S ( \rho ) = - k \trac (\rho \ln \rho) = -k
\lt \sqrt{\rho} | \sqrt{\rho} \ln \rho \rt \label{5} \ee

\noindent Depending on the context, the $ S $-functional
represents  the statistical uncertainty as to the actual classical
or quantum state of a system, the information carried by the
occurrence of one of the possible events (or the degree of
uncertainty as to which will be the next event), or the
thermodynamic entropy.

For each given set of values $ \langle H\rangle $, $\langle N_1
\rangle $, \dots, $\langle N_r \rangle $, in the range of the mean
value functionals (Eqs. \ref{4}) corresponding to the GM's, we
consider the subset of all elements $ \rho $ in $ \mathcal{P} $
that share the given mean values, i.e., \be \mathcal{P} _{\lgr
\langle H \rangle, \langle N_1 \rangle, \ldots , \langle N_r
\rangle \rgr} =  \lgr \rho \mbox{ in } \mathcal{P}\, |\, m ( \rho
; H ) = \langle H \rangle , m ( \rho ; N_i ) = \langle N_i \rangle
\mbox{ for } i = 1 , \ldots , r \rgr \label{6} \ee \noindent On
each such subset, i.e., for fixed mean values  $ \langle H\rangle
$, $\langle N_1 \rangle $, \dots, $\langle N_r \rangle $ of the
generators of the motion, the $ S $-functional (Eq. \ref{5})
achieves a unique maximum at the point \be \rho = e^{- \alpha}
\exp \Big( - \beta H + {\textstyle \sum_{i = 1}^r} \nu_i N_i \Big)
\quad \mbox{ where }\quad \alpha = \ln \trac \Big[\exp \Big( -
\beta H  + {\textstyle \sum_{i = 1}^r} \nu_i N_i \Big)\Big]
\label{7} \ee \noindent and, of course, $ \beta = \beta \lt
\langle H \rangle, \langle N_1 \rangle, \ldots , \langle N_r
\rangle \rt $ and $ \nu_i = \nu_i \lt \langle H \rangle, \langle
N_1 \rangle, \ldots , \langle N_r \rangle \rt$. \noindent It is
noteworthy that the maximum-$ S $ points satisfy the condition \be
\sqrt{\rho} \ln \rho = - a \sqrt{\rho} - b \sqrt{\rho} H +
{\textstyle \sum_{i = 1}^r} c_i \sqrt{\rho} N_i \label{11} \ee for
some real numbers $ a $, $b$ and $c_i $, $ i = 1 , \ldots , r $.
In words, the maximum-$ S $ element $ \rho $ is such that $
\sqrt{\rho} \ln \rho $ lies in the linear manifold generated by
elements $ \sqrt{\rho} $, $\sqrt{\rho} H $, $\sqrt{\rho} N_1 $,
\dots, $N_r $. Condition \ref{11} is satisfied not only by the
maximum-$ S $ elements given by Eq. \ref{7}, but also by the
elements given by \be \rho = e^{- a} B\,\exp \Big( - b H +
{\textstyle \sum_{i = 1}^r} c_i N_i \Big)B \quad \mbox{ where
}\quad a = \ln \trac \Big[ B\,\exp \Big( - b H  + {\textstyle
\sum_{i = 1}^r} c_i N_i \Big) B\Big] \label{12} \ee \noindent
where $ b = b \lt B ; \langle H \rangle, \langle N_1 \rangle,
\ldots , \langle N_r \rangle \rt $, $ c_i = c_i \lt B ; \langle H
\rangle, \langle N_1 \rangle, \ldots , \langle N_r \rangle \rt$
and $ B $ is any idempotent element in $ \mathcal{L} $ (i.e., $
B^2 = B $). Clearly, Eq. \ref{12} reduces to Eq. \ref{7} iff $ B =
I $ ($ I = $ constant function equal to 1 on the whole $ \Omega $
in CSM; $ I = \diag ( 1 ) $ in CIT; $I = $
 identity operator on $ \mathcal{H} $ in QSM and QT).

\section*{DYNAMICAL LAW DESIGN SPECIFICATIONS}

Our scope is to design a function $ F ( \cdot ) $ such that every
solution $ \rho ( t ) $ of the autonomous differential equation
\be \fr{d}{d t} \rho (t) = F ( \rho (t) ) \label{16} \ee \noindent
with $ \rho (0) $ anywhere in $ \mathcal{P} $
 satisfies the following conditions for all $t$'s, $-\infty< t <\infty
$:

\begin{description}
    \item[(i)] $ \rho (t) $ lies entirely in $ \mathcal{P} $ (no forward nor backward escape times);
    \item[(ii)] $ m ( \rho (t) ; H ) = m( \rho (0) ; H ) $, and $
    m ( \rho (t) ; N_1 ) = m ( \rho (0) ; N_i) $ for $ i = 1 ,
    \ldots , r $;
    \item[(iii)] $ S ( \rho (t+u) ) \geq S ( \rho (t) ) $ for all
    $ u > 0 $;
    \item[(iv)] within each subset
     $ \mathcal{P}_{ \lgr  \langle H \rangle, \langle N_1
\rangle, \ldots , \langle N_r \rangle \rgr} $,  the maximum-$ S $
element given by Eq. \ref{7}) is the
    only equilibrium solution that is stable according to
    Lyapunov \cite{Lyapunov}; all other equilibrium elements must not be stable.
\end{description}

Notice that requirement (iv) is most restrictive. For example,
within QSM, it rules out the von Neumann evolution equation $ ( F
(\rho) = -i ( H \rho - \rho H ) / \hbar ) $ because all the
stationary density matrices ( $ \rho $ such that $ H \rho = \rho H
$) are stable according to Lyapunov and, in general, there are
many more than a single one within each set $ \mathcal{P}_{ \lgr
\langle H \rangle, \langle N_1 \rangle, \ldots , \langle N_r
\rangle \rgr} $.

\section*{SOME NECESSARY MATHEMATICAL BACKGROUND}

Given a subset of elements $ A , B , \ldots , Z $ in $ \mathcal{L}
$, we denote by $ M ( A , B , \ldots , Z ) $ the Gram matrix \be
\lqu \begin{array}{cccc} (A|A) & (A|B) & \ldots & (A|Z) \\ (B|A) &
(B|B) & \ldots & (B|Z) \\ \vdots & \vdots & \ddots & \vdots \\
(Z|A) & (Z|B) & \ldots & (Z|Z) \end{array} \rqu \label{17} \ee
\noindent where $ ( \cdot | \cdot ) $ is the real symmetric inner
product defined on $ \mathcal{L} $. We denote by $ G ( A , B ,
\dots , Z ) $ the Gram determinant of $ A , B , \ldots , Z $ with
respect to  inner product $ ( \cdot | \cdot ) $, i.e., $ G ( A , B
, \ldots , Z ) = \det [ M ( A , B , \ldots , Z ) ] $. Matrix $ M (
A , B , \ldots , Z ) $ is nonnegative definite and $ G ( A , B ,
\ldots , Z ) $ is nonnegative. Elements $ A , B , \ldots , Z $ are
linearly independent (LI) iff their Gram determinant $ G ( A , B ,
\ldots , Z ) $ is nonzero and, hence, strictly positive.

Given a subset of elements $ A , B , \ldots , Z $ in $ \mathcal{L}
$, we denote by $ L ( A , B , \ldots , Z ) $ the linear manifold
spanned by all linear combinations with real coefficients of the
elements $ A , B , \ldots , Z $. With respect to the inner product
$ ( \cdot | \cdot ) $ defined on $ \mathcal{L} $, we denote the
projection of a given element $ V $ in $ \mathcal{L} $ onto a
linear manifold $ L $ by the symbol $ { (V) }_L $. $ { (V) }_L $
is the unique element in $ L $ such that $ ( { (V) }_L | X ) = ( V
| X ) $ for all $ X $ in $ L $.

The theory of Gram determinants, very seldom used in the physics
literature, offers a useful explicit way of writing the projection
$ { (V) }_L $ of $ V $ onto a given linear manifold $ L $. Let the
given linear manifold be $L= L ( A , B , \ldots , Z ) $, where
elements $ A , B , \ldots , Z $ need not be LI. Select any subset
of LI elements $ E_1 , E_2 , \ldots , E_m $ spanning $ L $, i.e.,
such that $ G ( E_1 , E_2 , \ldots , E_m ) > 0 $ and $ L ( E_1 ,
E_2 , \ldots , E_m ) = L $. By the definition of $ { (V) }_L $, $
( { (V) }_L | E_j ) = ( V | E_j ) $ for every $ j = 1 , 2 , \ldots
, m $, and $ { (V) }_L = \sum_{i = 1}^m v_i E_i $, where $ v_i $
are real scalars. Thus, \be {\textstyle \sum_{i = 1}^m} v_i ( E_i
| E_j ) = ( V | E_j ) \mbox{ for } j = 1 , 2 , \ldots , m
\label{18} \ee \noindent Because $ ( E_i | E_j ) = { [ M ( E_ 1,
E_2 , \ldots , E_m ) ] }_{i j} $ and the elements $ E_1 , E_2 ,
\ldots , E_m $ are LI, Eqs. \ref{18} are LI and can be solved for
the $ v_i $'s to yield \be v_i = {\textstyle \sum_{i = 1}^m} ( V |
E_j ) { [ M { ( E_1 , E_2 , \ldots , E_m ) }^{- 1} ] }_{j i} \quad
\mbox{for } i = i = 1 , 2 , \ldots , m \label{19} \ee \noindent
and, therefore, \be { (V) }_L = {\textstyle \sum_{i = 1}^m}
{\textstyle \sum_{j = 1}^m} ( V | E_j ) { [ M { ( E_1 , E_2 ,
\ldots , E_m ) }^{- 1} ] }_{j\, i} E_i \label{20} \ee

Alternatively,  Cramer's rule yields the  equivalent, more elegant
expression \be { (V) }_L = - \fr{1}{G ( E_1 , E_2 , \ldots , E_m
)} \det \lqu
\begin{array}{ccccc} 0 & E_1 & E_2 & \ldots & E_m \\ ( E_1 | V ) &
( E_1 | E_1 ) & ( E_1 | E_2 ) & \ldots & ( E_1 | E_m ) \\ ( E_2 |
V ) & ( E_2 | E_1 ) & ( E_2 | E_2 ) & \ldots & ( E_2 | E_m ) \\
\vdots & \vdots & \vdots & \ddots & \vdots \\ ( E_m | V ) & ( E_m
| E_1 ) & ( E_m | E_2 ) & \ldots & ( E_m | E_m ) \end{array} \rqu
\label{21} \ee

Below, for any given element $ \rho $ in the set $ \mathcal{P} $,
we will need to consider the projection of $ \sqrt{\rho} \ln \rho
$ onto the linear manifold $ L ( \sqrt{\rho} , \sqrt{\rho} H ,
\sqrt{\rho} N_1 , \ldots , \sqrt{\rho} N_r ) $ where $ \sqrt{\rho}
, \sqrt{\rho} H, \sqrt{\rho} N_1 , \ldots , \sqrt{\rho} N_r $ are
not necessarily LI. Using Eq. \ref{21} and Definitions \ref{1b},
\ref{1c} and \ref{1a} of the inner product $ ( \cdot | \cdot ) $,
we find

\be { ( \sqrt{\rho} \ln \rho ) }_{L ( \sqrt{\rho} , \sqrt{\rho} H
, \sqrt{\rho} N_1 , \ldots , \sqrt{\rho} N_r )} = - \fr{1}{G \lt
\sqrt{\rho} R_0 , \sqrt{\rho} R_1 , \ldots \sqrt{\rho} R_z \rt}
\times \ee \be \times \det \lqu \begin{array}{ccccc} 0 &
\sqrt{\rho} R_0 & \sqrt{\rho} R_1 & \ldots & \sqrt{\rho} R_z \\
\trac \rho \ln \rho & \trac \rho R_0^2 & \fr{1}{2} \trac \rho \lgr
R_0 , R_1 \rgr & \ldots & \fr{1}{2} \trac \rho \lgr R_0 , R_z \rgr
\\ \trac \rho R_1 \ln \rho & \fr{1}{2} \trac \rho \lgr R_1 , R_0
\rgr & \trac \rho R_1 & \ldots & \fr{1}{2} \trac \rho \lgr R_1 ,
R_z \rgr \\ \vdots & \vdots & \vdots & \ddots & \vdots \\ \trac
\rho R_z \ln \rho & \fr{1}{2} \trac \rho \lgr R_z , R_0 \rgr &
\fr{1}{2} \trac \rho \lgr R_z , R_1 \rgr & \ldots & \trac \rho
R_z^2 \end{array} \rqu \label{22} \ee

\noindent where $ \{ A , B \} = A B + B A $, and $ R_0 , R_1 ,
\ldots , R_z $ are a  subset of elements  in $ \mathcal{L} $ such
that $ L( \sqrt{\rho} R_0 , \sqrt{\rho} R_1 , \ldots , \sqrt{\rho}
R_z ) = L ( \sqrt{\rho} , \sqrt{\rho} H , \sqrt{\rho} N_1 , \ldots
, \sqrt{\rho} N_r ) $ and $ G ( \sqrt{\rho} R_0 , \sqrt{\rho} R_1
, \ldots , \sqrt{\rho} R_z ) > 0 $.

\section*{STEPEEST-$S$-ASCENT NONLINEAR EVOLUTION EQUATION}

With the above background,   the nonlinear evolution equation
proposed by the author
\cite{thesis,Frontiers,Cimento1,Cimento2,ArXiv1} to meet our
design specifications takes the compact form:

\be \fr{d}{d t} \rho (t) =  - \fr{i}{\hbar} [ H , \rho ] -
\fr{1}{2\tau} \lt \sqrt{\rho} D (\rho) + D^{\dag} (\rho)
\sqrt{\rho} \rt \label{23} \ee \be D (\rho) = \sqrt{\rho} \ln \rho
- { ( \sqrt{\rho} \ln \rho ) }_{L \lt \sqrt{\rho} , \sqrt{\rho} H
, \sqrt{\rho} N_1 , \ldots , \sqrt{\rho} N_r \rt} \label{24} \ee

\noindent where $ [ H , \rho] = H \rho - \rho H $ (= 0 within CSM
and CIT), $ \hbar $ is the reduced Planck constant (playing a role
only within QSM and QT), $ \tau $ is a characteristic time
constant, $ H , N_1 , \ldots , N_r $ are fixed GM's.

Notice  that $ D ( \rho (t) ) $ is orthogonal to the linear
manifold spanned by $ \sqrt{\rho} (t) , \sqrt{\rho} (t) H ,
\sqrt{\rho} (t) N_1 , \ldots , \sqrt{\rho} (t) N_r $ and the term
$ i [ H , \rho ] $ is orthogonal to  $ \sqrt{\rho} D (\rho) +
D^{\dag} (\rho) \sqrt{\rho} $.

Within QT, if the isolated and uncorrelated system $AB$ is
composed of subsystems $A$ (Alice) and $B$ (Bob) that are either
interacting ($H=H_\A\Otimes I_\B+I_\A\Otimes H_\B + V_{AB}$) or
correlated ($S\ne S_\A\Otimes I_\B+I_\A\Otimes S_\B $, where $
S=-\Boltz (P_{\Ran\, \rho}) \ln\rho$), or both, the proposed
equation takes the form \cite{Cimento2,ArXiv1} (we assume for
simplicity that $A$ and $B$ have no non-Hamiltonian GM's):
\begin{equation}{\displaystyle \ddt{\rho} = - \frac{i}{\hbar}}[H,\rho]
+ {\displaystyle  \frac{1}{ 2\Boltz\tau_\A} }\big(\sqA
D_\A\!+\!D_A^\dagger\sqA\big)\,\Otimes\, \rho_\B+ {\displaystyle
\frac{1}{ 2\Boltz\tau_\B} } \rho_\B\,\Otimes\,\big(\sqB
D_\B\!+\!D_B^\dagger\sqB\big)\nonumber\label{composite}\end{equation}
\begin{equation}\begin{tabular}{ll}
 $D_\A= \sqA (S)^\A-[\sqA (S)^\A]_{L\{\sqA,\sqA
(H)^\A\}}$ & $\qquad\qquad\qquad D_\B= \sqB (S)^\B-[\sqB
(S)^\B]_{L\{\sqB,\sqB (H)^\B\}}$  \\
  $(H)^\A=\Tr_\B
[(I_\A\Otimes \rho_\B) H]$ & $\qquad\qquad\qquad (H)^\B=\Tr_\A
[(\rho_\A\Otimes I_\B) H]$  \\
  $(S)^\A=\Tr_\B
[(I_\A\Otimes \rho_\B) S]$ &$\qquad\qquad\qquad (S)^\B=\Tr_\A
[(\rho_\A\Otimes I_\B) S]$
\end{tabular}\end{equation}
Despite the nonlinearity, the  structure of the non-Hamiltonian
terms in the equation prevents "no-signaling" violations.

All zero entropy states ($\rho^2=\rho$), even if $A$ and $B$ are
entangled, obey the Schroedinger equation ${\displaystyle
\ddt{\rho} = - \frac{i}{\hbar}}[H,\rho]$, thus not contradicting
any of the results of QM, although within QT, these solutions,
including the stationary states of QM ($\rho^2=\rho$, $\rho
H=H\rho$) are very weakly unstable limit cycles or equilibrium
states \cite{PRE}.

The proofs that this equations satisfy our  design specification
 are in Refs.
\cite{Cimento1,Cimento2,TwoLevel,Gheorghiu,ArXiv1}. Several other
intriguing features including Onsager's reciprocal relations are
discussed in Refs. \cite{Onsager,Gheorghiu,PRE}. Eq.
\ref{composite} generalizes  Eq. \ref{23} to composite systems
consistently with the additional design specs required to avoid
non-locality paradoxes \cite{Cimento2,ArXiv1}.

We believe that in view of its intriguing properties, this
evolution equation constitutes an important mathematical ``tool''
for a variety of non-equilibrium relaxation problems, not only
within our QT, but also within different contexts, such as CSM,
CIT, and QSM, as discussed here, as well as Quantum Information,
Biology, Sociology and Economics. Even in the Thermodynamics
context, where different schools of thought  notoriously have
contrasting perspectives on the physical meaning of entropy and
irreversibility, many important insights can nevertheless be
extracted from the richness of structure and the well-behaved and
self-consistent features of this relevant nonlinear dynamical
equation.

\bibliographystyle{unsrt}

\begin{thebibliography}{33}


\bibitem{MPLA1}G.P. Beretta, Mod. Phys. Lett. A {\bf 20}, 977 (2005).

\bibitem{HG}G.N. Hatsopoulos and E.P. Gyftopoulos, Found. Phys. {\bf 6}, 15, 127, 439, 561 (1976).

\bibitem{thesis}G.P. Beretta,  Sc.D. thesis, M.I.T.,
1981, unpublished, e-print quant-ph/0509116.


\bibitem{Frontiers}G.P. Beretta, in {\it Frontiers of
Nonequilibrium Statistical Physics,} NATO Advanced Study
Institute, Santa Fe, 1984, Eds. G.T. Moore and M.O. Scully (NATO
ASI Series B: Physics {\bf 135}, Plenum Press, NY, 1986), p. 193
and p. 205. G.P. Beretta, in {\it The Physics of Phase Space,}
Eds. Y.S. Kim and W.W. Zachary (Lecture Notes in Physics {\bf
278}, Springer-Verlag, NY, 1986), p. 441.


\bibitem{Cimento1}G.P. Beretta, E.P. Gyftopoulos, J.L. Park,
and G.N. Hatsopoulos, Nuovo Cimento B {\bf 82}, 169 (1984).

\bibitem{Cimento2}G.P.
Beretta, E.P. Gyftopoulos, and J.L. Park, Nuovo Cimento B {\bf
87}, 77 (1985). J. Maddox,
 Nature {\bf 316}, 11 (1985).


 \bibitem{Jaynes}E.T. Jaynes,Phys. Rev., Vol. 106,
  620 (1957); Vol. 108, 171 (1957).


\bibitem{TwoLevel}G.P. Beretta, Int. J. Theor. Phys. {\bf 24}, 119
(1985).


\bibitem{Lyapunov}G.P. Beretta, J. Math. Phys. {\bf 27}, 305 (1986).

\bibitem{Onsager}G.P. Beretta, Found. Phys. {\bf 17}, 365 (1987).

\bibitem{Gheorghiu}S.\
Gheorghiu-Svirschevski, Phys. Rev. A {\bf 63}, 022105, 054102
(2001).

\bibitem{ArXiv1}G.P. Beretta, e-print quant-ph/0112046.

\bibitem{PRE}G.P. Beretta, Phys. Rev. E {\bf 73}, 026113 (2006).






\bibitem{availableOnline}References
\cite{Frontiers,Cimento1,Cimento2,TwoLevel,Lyapunov,Onsager} are
available online at www.quantumthermodynamics.org.


\end{thebibliography}

    \end{document}